\documentclass{article}
\usepackage{spconf,amsmath,graphicx,hyperref,booktabs}
\usepackage{balance, multirow}

\title{Summary on The Multilingual Conversational Speech Language Model Challenge: Datasets, Tasks, Baselines, and Methods}
\name{
\parbox{\linewidth}{\centering
Bingshen Mu$^1$, Pengcheng Guo$^1$, Zhaokai Sun$^1$, Shuai Wang$^2$, Hexin Liu$^3$, Mingchen Shao$^1$,\\Lei Xie$^{1*}$, Eng Siong Chng$^3$, Longshuai Xiao$^4$, Qiangze Feng$^5$, Daliang Wang$^5$}\thanks{$^*$Corresponding author.}}
\address{$^1$Audio, Speech and Language Processing Group (ASLP@NPU), School of Computer Science, \\Northwestern Polytechnical University, Xi'an, China\\
$^2$School of Intelligence Science and Technology, Nanjing University\\
$^3$College of Computing and Data Science, Nanyang Technological University, Singapore\\
$^4$Huawei Technologies, China\\
$^5$Nexdata Technology Inc., USA
}

\begin{document}
\ninept
\maketitle
\begin{abstract}
\vspace{-4.5pt}
This paper summarizes the Interspeech2025 Multilingual Conversational Speech Language Model (MLC-SLM) challenge, which aims to advance the exploration of building effective multilingual conversational speech LLMs (SLLMs).
We provide a detailed description of the task settings for the MLC-SLM challenge, the released real-world multilingual conversational speech dataset totaling approximately 1,604 hours, and the baseline systems for participants.
The MLC-SLM challenge attracts 78 teams from 13 countries to participate, with 489 valid leaderboard results and 14 technical reports for the two tasks.
We distill valuable insights on building multilingual conversational SLLMs based on submissions from participants, aiming to contribute to the advancement of the community.
\end{abstract}
\vspace{-9pt}
\section{Introduction}
\vspace{-9pt}
\label{sec:intro}
Large language models (LLMs) have demonstrated remarkable capabilities across a wide range of natural language processing tasks, including text understanding, generation, and reasoning~\cite{grattafiori2024llama, team2024gemma, yang2024qwen25}.
These advancements have motivated efforts to extend LLMs beyond text, particularly to speech, which offers a more natural and direct form of communication.
Speech LLMs (SLLMs) denote LLMs that integrate the speech modality. Recent SLLMs have made significant progress in a wide range of tasks, including speech recognition~\cite{mu2024mmger, mu2025hdmole, mu2025mixture}, speech synthesis~\cite{anastassiou2024seed, wang2025spark}, speech enhancement~\cite{wang2024selm, kang2025llase}, and spoken dialogue systems~\cite{xu2025qwen2, wang2024freeze, defossez2024moshi}.

However, developing robust LLM-based spoken dialogue systems heavily relies on real-world conversational speech data, which embodies the complexities of human communication, including natural pauses, interruptions, speaker overlaps, and diverse conversational styles. The scarcity of such data, particularly in multilingual settings, poses a significant challenge to advancing in this field.


The Interspeech2025 Multilingual Conversational Speech Language Model (MLC-SLM) Challenge\footnote{https://www.nexdata.ai/competition/mlc-slm} aims to advance the exploration of building effective multilingual conversational SLLMs by releasing a real-world multilingual conversational speech dataset.
The challenge consists of two tasks: Task 1 focuses on multilingual conversational speech recognition, while Task 2 addresses multilingual conversational speech diarization and recognition.
The MLC-SLM challenge attracts 78 teams from 13 countries to participate, with 489 valid leaderboard results and 14 technical reports for the two tasks.
This paper intends to present the details of the dataset released for the MLC-SLM challenge, the setup and baseline systems for each task, and summarize the methods adopted by the participating teams, providing the community with insights into developing effective multilingual conversational SLLMs.

\vspace{-9pt}
\section{Task Settings and Rules}
\vspace{-9pt}
\label{sec:setting}
\textbf{Task 1: Multilingual Conversational Speech Recognition.} This task focuses on optimizing automatic speech recognition (ASR) accuracy in a multilingual conversational setting.
The objective of this task is to build a multilingual LLM-based ASR model.
Participants are provided with oracle segmentation and speaker labels for each recording.
We use word error rate (WER) and character error rate (CER) to evaluate the final results.
Specifically, we use CER to evaluate languages without clear word boundaries, including Japanese, Korean, and Thai.
Meanwhile, we use WER for the other languages.
Finally, we integrate the average recognition error rates of all languages through mixed error rate (MER).

\textbf{Task 2: Multilingual Conversational Speech Diarization and Recognition.} The objective of this task is to build a system for both speech diarization (SD) and recognition.
This task will provide no prior or oracle information during evaluation (e.g., no pre-segmented utterances or speaker labels).
We encourage pipeline-based or end-to-end systems, providing flexibility in design and implementation.
We use time-constrained minimum-permutation word error rate (tcpWER) and time-constrained minimum-permutation character error rate (tcpCER) to evaluate the final results.
Similar to Task 1, we use tcpCER to evaluate Japanese, Korean, and Thai, while tcpWER evaluates other languages.
The final ranking of all submissions in this task is based on time-constrained minimum-permutation mixed error rate (tcpMER).

\textbf{Rules.} All evaluation metrics are computed using the meeteval toolkit\footnote{https://github.com/fgnt/meeteval}.
Both tasks require participants to develop systems based on SLLMs.
Participants can use publicly available external datasets and pre-trained models, including speech foundation models and LLMs.
Data augmentation is allowed on the released training subset and may include, but is not limited to, the addition of noise or reverberation, speed perturbation, and tone modification.
Any use of the evaluation subsets for non-compliant purposes is strictly prohibited, including but not limited to using the evaluation subsets for model fine-tuning or training.
Participants cannot employ system fusion in either Task 1 or Task 2. Submitted results must be derived from a single system rather than through result fusion.

\begin{table}[t]
    \caption{Durations (hours) of the subsets in the MLC-SLM dataset.}
    \label{tab:table1}
    \centering
\scalebox{1}{
\begin{tabular}{lcccc}
\toprule
Language           & Train      & Dev      & Eval-1      & Eval-2      \\ \midrule
English-American   & 100.60     & 2.22     & 2.01        & 2.03        \\
English-Australian & 100.39     & 2.34     & 2.43        & 2.24        \\
English-British    & 100.48     & 2.23     & 2.03        & 2.15        \\
English-Filipino   & 100.36     & 2.09     & 2.02        & 2.06        \\
English-Indian     & 100.45     & 2.17     & 2.31        & 2.29        \\
French             & 100.38     & 2.26     & 2.07        & 2.11        \\
German             & 100.58     & 2.03     & 2.05        & 2.12        \\
Italian            & 100.67     & 2.10     & 2.31        & 2.25        \\
Japanese           & 100.44     & 2.08     & 2.19        & 2.18        \\
Korean             & 100.68     & 2.03     & 2.02        & 2.15        \\
Portuguese         & 100.33     & 2.18     & 2.14        & 2.07        \\
Russian            & 100.41     & 2.05     & 2.18        & 2.13        \\
Spanish            & 100.47     & 2.14     & 2.18        & 2.13        \\
Thai               & 100.50     & 2.12     & 2.17        & 2.26        \\
Vietnamese         & 100.45     & 2.15     & 2.08        & 1.99        \\ \midrule
Total              & 1507.22    & 32.18    & 32.19       & 32.22       \\ \bottomrule
\end{tabular}}
\vspace{-18pt}
\end{table}

\vspace{-9pt}
\section{Released Dataset}
\vspace{-9pt}
\label{sec:dataset}
Table~\ref{tab:table1} presents the statistics of the released MLC-SLM dataset. 
The dataset contains approximately 1604 hours of multilingual conversational data in total, including 1,507 hours for training (Train), 32 hours for development (Dev), 32 hours for Task 1 evaluation (Eval-1), and 32 hours for Task 2 evaluation (Eval-2).
Task 1 and Task 2 share the same Train and Dev subsets.
The Train and Dev subsets provide oracle segmentation, speaker labels, and transcriptions for building speech diarization and recognition systems.
The Eval-1 subset provides oracle segmentation and speaker labels, whereas the Eval-2 subset does not.

Each subset comprises 11 languages: English, French, German, Italian, Portuguese, Spanish, Japanese, Korean, Russian, Thai, and Vietnamese.
The English subsets contain accents from various regions: American, Australian, British, Filipino, and Indian.
Each recording contains a multi-turn conversational speech of around 20 minutes between two speakers on a randomly assigned topic, including celebrities, dreams, education, emotion, fashion, food, games, the Internet, movies, shopping, travel, etc.
The speakers of the conversations have diverse ages and genders.
The conversations are natural and fluent, with speakers engaging in meaningful exchanges around each topic.
All recordings are captured in quiet indoor environments using mobile phones such as iPhones, with a sampling rate of 16kHz.
This dataset provides a rich resource for building multilingual conversational SLLMs, addressing the challenges of linguistic diversity, speaker variability, and contextual understanding.

In addition, the speech data and corresponding segmentation, speaker labels, and transcriptions for the Train and Dev subsets have been released to participants, along with detailed metadata that includes conversation topics, speaker information, recording conditions, and speech data encoding information.
The speech data, segmentation, and speaker labels for the Eval-1 subset and the speech data for the Eval-2 subset have also been released to participants.
After the challenge, we release the segmentation, speaker labels, and transcriptions for both the Eval-1 and Eval-2 subsets to enable wider application of the MLC-SLM dataset and promote further research: \textit{https://huggingface.co/datasets/bsmu/MLC-SLM-Eval}.

\begin{figure}[t]
\centering
\includegraphics[width=1.0\columnwidth]{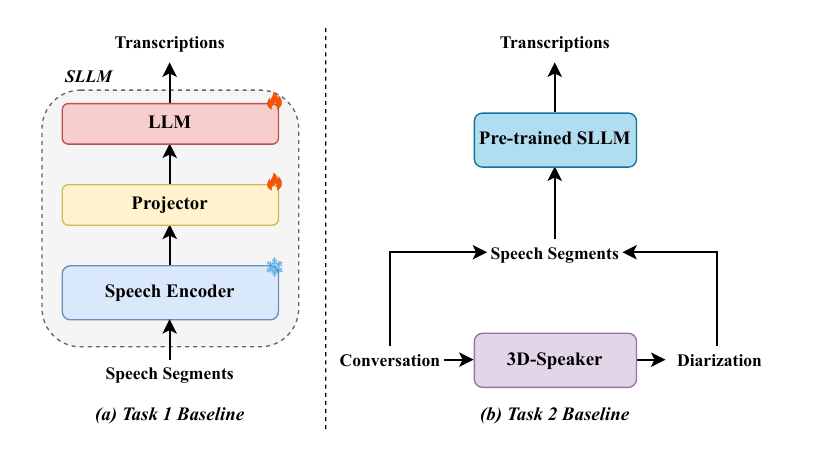}
\caption{Architectures of the baseline systems for Task 1 and Task 2.}
\vspace{-9pt}
\label{fig1}
\end{figure}

\begin{table}[t]
    \caption{Results of the baseline systems on the Dev and Eval subsets. Whisper refers to the vanilla Whisper-large-v3, B-Q refers to the SLLM with Qwen2.5-7B, B-L refers to the SLLM with Llama3.1-8B, SD refers to SD with overlap detection, while OSD refers to SD without overlap detection.}
    \label{tab:table2}
    \centering
\scalebox{0.97}{
\begin{tabular}{lccccc}
\toprule
\multirow{2}{*}{Subset}        & \multicolumn{3}{c}{Task 1 MER (\%) $\downarrow$} & \multicolumn{2}{c}{Task 2 tcpMER (\%) $\downarrow$} \\ \cmidrule{2-6} 
       & Whisper     & B-Q       & B-L       & SD \& B-L         & OSD \& B-L         \\ \midrule
Dev    & 16.82       & 21.49     & 21.56     & 76.12             & 81.85              \\
Eval-1 & 17.33       & 20.21     & 20.17     & -                 & -                  \\
Eval-2 & -           & -         & -         & 60.39             & 65.07              \\ \bottomrule
\end{tabular}}
\vspace{-18pt}
\end{table}

\begin{table*}[t]
    \caption{Results of the top ranking teams with a valid technical report on the Eval-1 subset in Task 1 and their main techniques.}
    \label{tab:table3}
    \centering
\scalebox{0.63}{
\begin{tabular}{lcccccccc}
\toprule
\multirow{2}{*}{Team} & \multicolumn{3}{c}{SLLM Architecture}                                                                                                                                       & \multirow{2}{*}{Training Strategy} & \multirow{2}{*}{Multilingual Adaptation}                                                   & \multirow{2}{*}{Data Augmentation}                                                                         & \multirow{2}{*}{External Data~(h)} & \multirow{2}{*}{MER (\%) $\downarrow$} \\ \cmidrule{2-4}
                      & Speech Encoder                                                         & Projector                                                                        & LLM Decoder         &                                    &                                                                                            &                                                                                                            &                                &                           \\ \midrule
TENP~\cite{xue2025tea}                  & \begin{tabular}[c]{@{}c@{}}Whisper-large-v3\\ MMS-1B\end{tabular}      & Language-adapted connector                                                       & Qwen3-8B-Base       & Two-stage                          & \begin{tabular}[c]{@{}c@{}}MoE LoRA\\ Language-adapted fusion\end{tabular}                 & \begin{tabular}[c]{@{}c@{}}Low-quality speech filter\\ Spectral enhancement\\ Speed variation\end{tabular} & $\sim$179k                    & 9.60                      \\ \midrule
Sixteen-years~\cite{gao2025triple}         & Whisper-large-v3                                                       & \begin{tabular}[c]{@{}c@{}}Downsampling module\\ Two-layer linear\end{tabular}   & Qwen3-8B-Base       & Three-stage                        & -                                                                                          & \begin{tabular}[c]{@{}c@{}}SpecAugment\\ Speed perturbation\end{tabular}                                   & $\sim$30k                     & 9.67                      \\ \midrule
T-ASR~\cite{li2025transsion}                 & Whisper-large-v3                                                       & \begin{tabular}[c]{@{}c@{}}Downsampling module\\ Two-layer linear\end{tabular}   & Qwen2.5-7B-Instruct & One-stage                          & -                                                                                          & -                                                                                                          & $\sim$7k                      & 9.83                      \\ \midrule
MegaAIS~\cite{meng2025ilt}               & \multicolumn{3}{c}{Qwen2-Audio}                                                                                                                                                 & Three-stage                        & -                                                                                          & \begin{tabular}[c]{@{}c@{}}Text augmentation\\ Audio augmentation\end{tabular}                             & $\sim$42k                      & 10.08                     \\ \midrule
NTU-Speechlab\cite{peng2025ntu}         & Whisper-large-v3                                                       & Two-layer linear                                                                 & Gemma-2-2B          & One-stage                          & Language specific prompt                                                                   & \begin{tabular}[c]{@{}c@{}}Speed perturbation\\ Volume perturbation\end{tabular}                           & $\sim$16k                      & 10.58                     \\ \midrule
Seewo~\cite{li2025seewo}                 & Whisper-large-v3-turbo                                                 & \begin{tabular}[c]{@{}c@{}}One-layer convolution\\ Two-layer linear\end{tabular} & Babel-9B-Chat       & Multi-stage                        & \begin{tabular}[c]{@{}c@{}}Language specific token\\ Language specific prompt\end{tabular} & \begin{tabular}[c]{@{}c@{}}Additive noise\\ Reverberation\end{tabular}                                     & -                              & 11.57                     \\ \midrule
May~\cite{mei2025shnu}                   & \begin{tabular}[c]{@{}c@{}}Whisper-large-v3\\ mHuBERT-147\end{tabular} & \begin{tabular}[c]{@{}c@{}}Two-layer convolution\\ Two-layer linear\end{tabular} & Qwen2.5-7B          & Three-stage                        & Language specific prompt                                                                   & -                                                                                                          & -                              & 11.76                     \\ \midrule
Eloquence~\cite{concina2025eloquence}                   & Whisper-large-v3-turbo & \begin{tabular}[c]{@{}c@{}}Downsampling module\\ One-layer linear\end{tabular} & EuroLLM-1.7B          & One-stage                        & -                                                                   & -                                                                                                          & -                              & 15.15                     \\ \midrule
HighAccuracy~\cite{nguyen2025qwen}                   & Whisper-large-v3 & Two-layer linear & Gemma-3-12B          & Three-stage                        & -                                                                   & SpecAugment                                                                                                          & -                              & 16.63                     \\ \bottomrule
\end{tabular}}
\vspace{-18pt}
\end{table*}

\vspace{-9pt}
\section{Baselines}
\vspace{-9pt}
\label{sec:baselines}
We release the code for data processing, model training, and model inference of the baseline systems for both tasks to facilitate quick start-up and reproducible research for the participants\footnote{https://github.com/mubingshen/MLC-SLM-Baseline}.
Figure~\ref{fig1} illustrates the architectures of the baseline systems for both tasks.

The Task 1 baseline system adopts a standard SLLM paradigm, which consists of a speech encoder, a projector layer, and a LLM decoder.
Specifically, the speech encoder uses the vanilla Whisper-large-v3\footnote{https://huggingface.co/openai/whisper-large-v3} encoder, the projector layer consists of two convolutional layers with GELU activation, two linear layers with ReLU activation, and LayerNorm, and the LLM uses Qwen2.5-7B\footnote{https://huggingface.co/Qwen/Qwen2.5-7B} and Llama3.1-8B\footnote{https://huggingface.co/meta-llama/Llama-3.1-8B}.
We enable the SLLM to complete the ASR task through a two-stage training strategy.
In the first stage, we only train the projector layer.
In the second stage, we load the projector layer parameters from the first stage and jointly train the projector layer and the LLM using Low-Rank Adaptation (LoRA)~\cite{hu2022lora}.

The Task 2 baseline system follows a cascade pipeline.
Specifically, we use the 3D-Speaker toolkit\footnote{https://github.com/modelscope/3D-Speaker} for SD and the Task 1 pre-trained SLLM model for ASR.
We obtain the results for Task 2 through a three-stage strategy.
In the first stage, we fine-tune the pyannote-segmentation module\footnote{https://huggingface.co/pyannote/segmentation-3.0} for overlap detection.
In the second stage, we load the segmentation module from the first stage and perform inference on the raw conversation in the Dev and Eval-2 subsets to obtain SD results.
In the third stage, we split the raw conversation based on the SD results and input the splited speech segments into the Task 1 pre-trained SLLM model to generate transcriptions.

Table~\ref{tab:table2} presents the results of the baseline systems on the Dev and Eval subsets.
For Task 1, the vanilla Whisper-large-v3 outperforms the SLLMs with Qwen2.5-7B or Llama3.1-8B. We attribute this to two factors: first, the baseline SLLMs do not utilize language identification, whereas it is used when inferring with the vanilla Whisper-large-v3; second, SLLMs exhibit severe hallucination issues during inference.
Moreover, only a minimal performance difference between Qwen2.5-7B and Llama3.1-8B.
For Task 2, the SD module without overlap detection performs better because the speaker overlap in the dataset is too brief to be considered significant.
Furthermore, due to the large number of short speech segments resulting from the baseline SD-based segmentation, the pre-trained SLLM from Task 1 exhibits a high tcpMER.

\vspace{-9pt}
\section{Methods of Task 1}
\vspace{-9pt}
\label{sec:methods1}
A total of 24 teams submitted their results on the Eval-1 subset for Task 1, and the MER results and main techniques used by the top 9 ranking teams that submitted technical reports are listed in Table~\ref{tab:table3}.
We summarize insights on how multilingual conversational SLLMs achieve superior ASR performance, covering aspects such as SLLM architectures, training strategies, multilingual adaptation, data augmentation, and conversation context utilization.
\vspace{-9pt}
\subsection{SLLM Architecture}
\vspace{-4.5pt}
Most teams build their SLLMs using a speech encoder, a projector, and an LLM decoder, similar to the architecture shown in Figure~\ref{fig1}.
The Whisper~\cite{radford2023robust} family is the most common source for speech encoders due to its powerful and extensive multilingual ASR capabilities, with Whisper-large-v3 being the most frequently used encoder.
Some teams also use Whisper-large-v3-turbo\footnote{https://huggingface.co/openai/whisper-large-v3-turbo} as the speech encoder, but its performance is slightly inferior to that of the Whisper-large-v3 encoder~\cite{li2025transsion, li2025seewo, concina2025eloquence, wang2025rag}.
In addition, two teams adopt dual-encoder architectures, employing encoders trained with weak-supervised and self-supervised learning, respectively, to provide more comprehensive speech features~\cite{xue2025tea, mei2025shnu}.
The projector functions to downsample speech features and align them with the LLM text embedding space. Downsampling is commonly achieved by concatenating adjacent speech features or using convolutional layers, while linear layers are applied to map the downsampled speech features into the LLM text embedding space.
As an alternative downsampling method in the projector, Q-former~\cite{li2023blip} transforms speech features into learnable queries of a specified length; however, its performance is unsatisfactory~\cite{concina2025eloquence}.
A specially designed language-adapted projector performs weighted fusion of the dual-encoder outputs conditioned on language identification, and the fused embeddings are input into the LLM while generating CTC prompts~\cite{xue2025tea}.
The LLM decoders come from diverse sources, including the Llama-3.2\cite{grattafiori2024llama}, Gemma-2~\cite{team2024gemma}, Gemma-3~\cite{team2025gemma}, Qwen-2.5~\cite{yang2024qwen25}, Qwen-3~\cite{yang2025qwen3}, Babel~\cite{zhao2025babel}, EuroLLM~\cite{martins2025eurollm} family, etc.
We find that different LLM decoders exhibit no significant performance gap within SLLMs.
Furthermore, fine-tuning existing SLLMs, such as Qwen2-Audio\footnote{https://huggingface.co/Qwen/Qwen2-Audio-7B} and Phi-4-multimodal-instruct\footnote{https://huggingface.co/microsoft/Phi-4-multimodal-instruct}, is also a feasible solution~\cite{meng2025ilt, saengthong2025unified}.
\vspace{-9pt}
\subsection{Training and Inference Strategy}
\vspace{-4.5pt}
Strategies for training SLLMs for the ASR task are diverse, including one-stage, two-stage, three-stage, and multi-stage training.
One-stage training involves simultaneously training the projector and the LLM, optionally applying LoRA for parameter-efficient fine-tuning~\cite{li2025transsion, peng2025ntu, concina2025eloquence, peng2025bi, lin2025dku}.
Two-stage training includes a monolingual-to-multilingual strategy~\cite{saengthong2025unified}, a strategy that first trains the speech encoder and projector before jointly training the projector and LLM~\cite{xue2025tea}, and the same training strategy as the baseline described in Section~\ref{sec:baselines}.
Each stage of the three-stage training involves optimizing one or several components among the speech encoder, projector, and LLM decoder. For example, train the three components separately, or jointly train the speech encoder and projector, the projector and LLM, or all three components together. The ultimate goal is to optimize all three components to better adapt to the ASR task~\cite{gao2025triple, mei2025shnu, nguyen2025qwen}.
For three-stage training that fine-tunes an existing SLLM, different data types are used in each training stage to enhance its ASR performance~\cite{meng2025ilt}.
Multi-stage training further exploits the potential of SLLMs through chain-of-thought reasoning and various reinforcement learning strategies~\cite{li2025seewo}.
During inference, model averaging remains a key method for improving the performance of SLLMs on the ASR task during inference~\cite{xue2025tea, li2025transsion, peng2025ntu}.
Moreover, some LLM inference parameters are crucial for SLLM inference in the ASR task.
For example, the optimal beam size for different SLLMs may vary when inferring on different languages, and whether the generation process samples the next token also impacts ASR performance~\cite{gao2025triple, meng2025ilt, peng2025ntu, li2025seewo}.
\vspace{-9pt}
\subsection{Multilingual Adaptation}
\vspace{-4.5pt}
Multilingual adaptation enables SLLMs to generate outputs that better match the input speech language, mitigating the issue of language confusion in the output.
Language-specific prompts offer the most straightforward way to achieve multilingual adaptation by translating the prompt ``Please transcribe the speech" into the same language as the input speech~\cite{peng2025ntu, mei2025shnu, peng2025bi}.
Combining language-specific trainable special tokens with language-specific prompts can further enhance multilingual adaptation~\cite{li2025seewo}.
Furthermore, training a separate set of the projector and LLM LoRA parameters for each language is another method to achieve multilingual adaptation~\cite{lin2025dku}.
Going further, applying MoE LoRA in dual encoders and performing weighted fusion of speech features conditioned on language identification allows both the fused speech features and their generated CTC prompts to serve as another form of language-specific prompts, thereby achieving multilingual adaptation that, to some extent, combines all the previously discussed methods~\cite{xue2025tea}.
\vspace{-9pt}
\subsection{Data Augmentation}
\vspace{-4.5pt}
Since the released MLC-SLM dataset is relatively small in scale, many teams applied various data augmentation techniques.
Given that publicly available external datasets are permitted, this became one of the key methods for data augmentation. 
However, as some external multilingual datasets are of low quality, it is necessary to filter and remove them.
Furthermore, since the scale of publicly available speech data varies greatly across languages, a data balancing strategy is employed to enhance performance on lower-resource languages~\cite{xue2025tea}.
Common data augmentation techniques, such as SpecAug, SpecSubstitute, speed perturbation, additive noise, and reverberation simulation, are widely used~\cite{xue2025tea, gao2025triple, li2025seewo, concina2025eloquence, saengthong2025unified, nguyen2025qwen}.
More complex text-and-speech combined augmentation techniques can further enhance SLLM performance~\cite{meng2025ilt}.
\vspace{-9pt}
\subsection{Conversation Context Utilization}
\vspace{-4.5pt}
The high degree of contextual relevance in conversational speech presents an excellent potential for conversational SLLMs that utilize context.
Incorporating a fixed number of adjacent historical or future utterance hypotheses as conversation context can lead to SLLM performance that surpasses SLLMs trained with a large amount of extra data.
Moreover, using ground truth transcriptions of historical or future utterances can further explore the potential of conversation context utilization~\cite{li2025seewo, peng2025bi, mu2025hearing}.
Another method enhances the SLLM’s ability to capture the semantic coherence between the current utterance content and its conversation context by contrastive learning, which aligns the current utterance with its associated context~\cite{concina2025eloquence}.
\vspace{-9pt}
\section{Methods of Task 2}
\vspace{-9pt}
A total of 10 teams submitted their results on the Eval-2 subset for Task 2, and the tcpMER results and main techniques used by the top 6 ranking teams that submitted technical reports are listed in Table~\ref{tab:table4}.
We summarize insights on how multilingual conversational SLLMs can combine SD and ASR to achieve superior performance.
\vspace{-9pt}
\subsection{Cascade Pipeline}
\vspace{-4.5pt}
The cascade pipeline is the most common and widely used method for SD and ASR, comprising three key components: (1) a voice activity detection (VAD) module for accurate speech segment timestamp prediction; (2) a speaker embedding model for discriminating speaker representations; and (3) a speaker clustering module for grouping speakers.
Most teams use the same FSMN-based VAD module\footnote{https://modelscope.cn/models/iic/speech\_fsmn\_vad\_zh-cn-16k-common-pytorch} as the baseline system~\cite{meng2025ilt, xue2025tea, li2025seewo, gao2025triple}, while one employs the innovative Sequence-to-Sequence Neural Diarization (S2SND)~\cite{cheng2025sequence} model as their VAD module~\cite{lin2025dku}.
The speaker embedding module has multiple options, including CAM++\footnote{https://modelscope.cn/models/iic/speech\_campplus\_sv\_zh\_en\_16k-common\_advanced}, ERes2Net-large\footnote{https://modelscope.cn/models/iic/speech\_eres2net\_large\_200k\_sv\_zh-cn\_16k-common}, and ResNet~\cite{he2016deep} models.
Post-processing of SD results can significantly improve ASR performance, including setting a maximum speech segment duration~\cite{li2025seewo, meng2025ilt} and concatenating adjacent segments from the same speaker~\cite{xue2025tea}.
After obtaining the SD results, the cascade pipeline methods split the raw conversational speech, then apply the pre-trained SLLMs from Task 1 to infer the speech segments for transcriptions.
Notably, the best-performing SLLM in Task 1 does not necessarily perform optimally when inferring speech segments obtained through SD results for transcriptions, due to the limited generalization ability of the best-performing SLLM concerning varying segment lengths from SD results~\cite{meng2025ilt}.
\vspace{-9pt}
\subsection{Semi-Combination}
\vspace{-4.5pt}
A deeper combination of SD and ASR is imperative to reduce the complexity of cascade pipelines.
One semi-integrated method takes the raw conversational speech with the corresponding set of speaker-enrollment triplets—including speaker embeddings and start–end timestamps—as input to a diarization-aware SLLM, which then generates transcriptions for the speech segments represented by each triplet~\cite{lin2025dku}.
Another semi-integrated method first employs an innovative DiariZen~\cite{han2025leveraging} as the SD model to generate frame-level Silence-Target-NonTarget-Overlap (STNO) masks for each speaker in the raw conversational speech, and then feeds these masks into an enhanced Whisper model DiCoW~\cite{polok2026dicow} to produce the transcription of the raw conversation~\cite{polok2025but}.
\vspace{-9pt}
\subsection{End-to-End}
\vspace{-4.5pt}
The end-to-end method is to employ a single SLLM to jointly perform SD and ASR, where the model predicts speaker labels, segment boundaries (start and end timestamps), and the corresponding transcriptions directly from the raw conversation speech.
Only one team explores the end-to-end method, which combines a local sliding non-overlapping inference window with prompt-based context from previous speaker turns to predict who is speaking and what is being said.
The end-to-end SLLM is trained with a Local Diarization and Recognition format to learn structured speaker–text alignments. During inference, it leverages updated speaker context and predicted speaker prompts to process the raw conversation speech within local windows, ensuring coherence iteratively~\cite{saengthong2025unified}.
\begin{table}[t]
    \caption{Results of the top ranking teams with a valid technical report on the Eval-2 subset in Task 2 and their main techniques.}
    \label{tab:table4}
    \centering
\scalebox{0.70}{
\begin{tabular}{lcccc}
\toprule
\multirow{2}{*}{Team} & \multicolumn{2}{c}{SD} & \multirow{2}{*}{SD-ASR Type} & \multirow{2}{*}{tcpMER (\%) $\downarrow$} \\ \cmidrule{2-3}
                      & VAD   & Spk. Embed.    &                                   &                              \\ \midrule
MegaAIS~\cite{meng2025ilt}               & FSMN  & CAM++          & Cascade pipeline                  & 16.53                        \\
TENP~\cite{xue2025tea}                  & FSMN  & ERes2Net-large & Cascade pipeline                  & 17.49                        \\
Seewo~\cite{li2025seewo}                 & FSMN  & ResNet-101     & Cascade pipeline                  & 17.67                        \\
DKU~\cite{lin2025dku}                   & S2SND & ResNet-34      & Semi-combination                  & 18.08                        \\
Sixteen-years~\cite{gao2025triple}         & FSMN  & CAM++          & Cascade pipeline                  & 19.27                        \\
ST-ShinozakiLab~\cite{saengthong2025unified}       & -     & -              & End-to-end                        & 27.25                        \\ \bottomrule
\end{tabular}}
\vspace{-18pt}
\end{table}
\vspace{-9pt}
\section{Conclusion}
\vspace{-9pt}
This paper summarizes the Interspeech2025 MLC-SLM challenge, which aims to advance the exploration of building effective multilingual conversational SLLMs.
After distilling all submissions, we observe that multilingual conversational SLLMs have already established diverse and effective solutions in model architecture, training strategies, and data augmentation.
However, the methods for multilingual adaptation, conversation context utilization, and the processing of speaker and temporal information for speech diarization remain overly straightforward and have not yielded the expected benefits, indicating further exploration.
\clearpage
\balance
\bibliographystyle{IEEEbib}
\bibliography{refs}

\end{document}